\documentclass{iau_FM}
\usepackage{graphicx}
\usepackage{url}

\title[A paradigm to develop new contributors to Astronomy] 
{A paradigm to develop \\new contributors to Astronomy}

\author[Maravelias et al.]   
{Grigoris Maravelias$^{1,2,3}$\thanks{contact: gmaravel@physics.uoc.gr}, Emmanouel Vourliotis$^{1,4}$, Krinio Marouda$^1$, Ioannis Belias$^1$, Emmanouel Kardasis$^1$, Pierros Papadeas$^1$, Iakovos D. Strikis$^1$, Eleftherios Vakalopoulos$^1$, Orfefs Voutyras$^5$}

\affiliation{$^1$ Hellenic Amateur Astronomy Association, Athens, Greece, $^2$ Foundation for Research and Technology-Hellas, Greece, $^3$ University of Crete, Heraklion, Greece, $^4$ National and Kapodistrian University of Athens, Zografou, Greece, $^5$ National Technical University of Athens, Zografou, Greece}

\pubyear{2015}
\jname{Astronomy in Focus, Volume 1} 
\editors{Piero Benvenuti, ed.}
\begin{document}

\maketitle

\begin{abstract}

One of the most regular activities of amateur clubs is scientific outreach, a paramount channel to disseminate scientific results. It is typically performed through talks given by both experts (professional astronomers) and non-experts to a diverse audience, including amateur astronomers. However, this is a rather passive, one-way, approach. The advance of technology has provided all the tools that can help the audience/amateurs to become more active in the scientific output. What is often missing is the proper guidance. To address that within the Greek amateur community the Hellenic Amateur Astronomy Association materialized a training program (free-of-charge and open-accessed) to develop scientific thought and the practical capabilities for amateurs to produce valuable results. The program ran from November 2014 to May 2015 focusing each session (month) to: the Sun, variable stars, comets, planets, artificial satellites, meteors. A professional and/or an experienced amateur astronomer was leading each session consisting of a theoretical part (highlights of the field, necessary observational techniques) and a hands-on part (observations and data analysis). At least 50 unique participants gained significant experience by following parts or the complete program.

\keywords{sociology of astronomy, miscellaneous: professional-amateur collaborations}
\end{abstract}

\firstsection 
\section{Introduction}

Nowadays in Astronomy, the advance of technology, both in hardware and software level, offers the appropriate tools for every interesting individual to participate in scientific output (e.g. \cite{Lintott2008, Mousis2014}). However, a successful outcome is not guaranteed unless proper training and guidance is provided. Driven by this lack, the Hellenic Amateur Astronomy Association took an initiative to materialize a free-of-charge and open-accessed training program to develop scientific thought and the practical capabilities for amateurs to produce scientifically valuable results.

\section{Methodology}
The program ran from November 2014 through May 2015, as a series of Observational Astronomy "months". Each month focused on a different field, starting with the Sun and continuing with variable stars, comets, planets of the Solar system, artificial objects (satellites), and meteors. A professional astronomer or an amateur with extensive experience in the field was responsible to lead each month and organize the necessary material to be used (building upon our previous experience, e.g. \cite{Voutyras2013}). Each session composed of a series of meetings (on weekends) split in two parts.

Initially, the participants were introduced to the field. An overview of the amateur contributions was presented to trigger their interest. Then, a detailed description of the required observational techniques for the field followed, e.g. visual/CCD photometry of variable stars. For this, we followed the best practices as determined by main international professional-amateur organizations.

The second part of each focused "month" was devoted to hands-on experience, including a number of exercises and homework (at later stages), such as finding the Wolf number for the Sun, performing photometry on pre-defined data, etc. Then, using the methodology presented at the first part they obtained their own observations, either through their own equipment or setups offered by the organizers. This step was followed by proper reduction of the acquired data (e.g. bias and flat-field corrections for photometry). The last step was to obtain the final product that was subsequently submitted to the various international databases.

\section{Results and Impact}
In total, we had about 100 non-unique participations (ranging from 5 to 50 per month) throughout the whole series of these courses. However, not all participants completed them successfully, i.e. follow all talks, work on the exercises, and finish the projects. In the case of planetary imaging (\cite{Kardasis2015}) 11 out of 48 participants finally managed to reach the final stage. This is an important figure when compared to the number of active Greek amateurs (a few hundreds). Despite that, the dropout rate from the program at $\sim77\%$ requires further investigation. A number of notable reasons is: (i) the content may have been too elaborated, as not all participants shared the same knowledge, (ii) difficulties to apply theory into practice (possibly some people were not interested finally in observations), (iii) for the months focusing on the variable stars and meteors, the lead coordinator was involved remotely, which led to a poorer contact with the participants, (iv) timing constraints (successive weekends) hindered the commitment of some participants, (v) last but not least, it is possible that different expectations of the participants were not met. 

The main goal of this initiative was to introduce interested amateurs to the scientific approach of Observational Astronomy. With this activity we managed to reach a significant number of people. Following an open-access approach, all the material used has been archived to our website\footnote{\url{http://www.hellas-astro.gr} currently in Greek, as English-speaking individuals do have many other sources to look into.}, so that this work can be used by any interested individual or group, and become a long-term guide. Currently (2018), a few years after the completion of this project, we notice that new contributors do show up, although their absolute number is rather slim.\\ 

Acknowledgments: GM acknowledges IAU travel fund.

\end{document}